\documentclass[10pt, conference, compsocconf]{IEEEtran}
\usepackage{amssymb,amsmath,amsfonts}
\usepackage{times}
\usepackage{dsfont}
\usepackage{stmaryrd}
\usepackage{textcomp}
\usepackage{graphicx}
\usepackage{cite}

\usepackage{stfloats}
\newtheorem{definition}{Definition}[section]

\begin{document}

\title{A Security Framework for Wireless Sensor Networks: Theory and Practice}

\author{\IEEEauthorblockN{Christophe Guyeux, Abdallah Makhoul and Jacques M. Bahi}
\IEEEauthorblockA{Femto-ST Institute\\
 Computer science departement (DISC)\\
University of Franche-Comt\'e\\
Rue Engel-Gros, 90016 Belfort, France\\
\{christophe.guyeux, abdallah.makhoul, jacques.bahi\}@univ-fcomte.fr}
}

%\IEEEoverridecommandlockouts
%\IEEEpubid{\makebox[\columnwidth]{978-1-4673-2480-9/13/\$31.00 ~\copyright~2013 IEEE \hfill} \hspace{\columnsep}\makebox[\columnwidth]{ }}

\maketitle 
\begin{abstract}
Wireless sensor networks are often deployed in public or otherwise untrusted
and even hostile environments, which prompts a number of security issues. Although security is a necessity
in other types of networks, it is much more so in sensor networks due to the resource-constraint, susceptibility to physical capture, and wireless nature. In this work we emphasize two security issues: (1) secure communication infrastructure and (2) secure nodes scheduling algorithm. Due to resource constraints, specific
strategies are often necessary to preserve the network's lifetime and its quality of
service. For instance, to reduce communication costs nodes can go to sleep mode periodically (nodes scheduling). These strategies must be proven as secure, but protocols used to guarantee this security must be compatible with the resource preservation 
requirement. To achieve this goal, secure communications in such networks will
be defined, together with the notions of secure scheduling. Finally, some of these security properties 
will be evaluated in concrete case studies.
\end{abstract}

%%%%%%%%%%%%%%%%%%%%%%%%%%%%%%%%%%%%%%%%%%%%%%%%%%%%%%%%%%%%%%%%%%%%%%%%%%%%%%%%%%%%%%%%%%%%%%%%
%%%%%%%%%%%%%%%%%%%%%%%%%%%%%%%%%%%%%%%%%%%%%%%%%%%%%%%%%%%%%%%%%%%%%%%%%%%%%%%%%%%%%%%%%%%%%%%%%

%\author[Bahi]{Jacques M. Bahi}
%\author[Bahi]{Christophe Guyeux}
%\author[Bahi,Guyeux]{Abdallah Makhoul}

%\address[Bahi]{FEMTO-ST Institute, UMR 6174 CNRS\\
%University of Franche-Comt\'{e}\\
%16, route de Gray\\
%25000 Besan\c con, France\\
%\{jacques.bahi, christophe.guyeux, abdallah.makhoul\}@femto-st.fr}
%\address[Guyeux]{Authors in alphabetic order}

\section{Introduction}

In the last few years, wireless sensor networks (WSN) have gained increasing attention from both the research community
and actual users. As sensor nodes are generally battery-powered devices, the critical aspects to face concern how to reduce
the energy consumption of nodes, so that the network lifetime can be extended to reasonable times. Therefore, energy conservation is a key issue in the design of systems based on wireless sensor networks. In the literature, we can find different techniques to extend the sensor network lifetime~\cite{87}. For example, energy
efficient protocols are aimed at minimizing the energy consumption during network activities. However, a large amount of energy is consumed by node components (CPU, radio, etc.) even if they are idle. Power management schemes are thus used for switching off node components that are not temporarily needed~\cite{47,46}. Other techniques suitable to reduce the energy consumption of sensors is data acquisition (i.e. sampling or transmitting) reduction as data fusion and aggregation~\cite{88,89}.

On the other hand, sensor networks are often deployed in unattended even hostile environments, thus leaving these networks vulnerable to passive and active attacks by the adversary. The communication between sensor nodes can be eavesdropped by the
adversary and can forge the data. Sensor nodes should be resilient to these attacks. Therefore, One of the major challenges in such networks is how to provide connection between sensors and the base station and how to exchange the data while maintaining the security requirements and taking into consideration their limited resources. In this paper we emphasize two security issues:

{\bf Secure communication infrastructure:} In wireless sensor networks a sensor node generally
senses the data and sends to its neighbor nodes or to the sink. Stationary adversaries
equipped with powerful computers and communication
devices may access whole WSN from a remote location. For instance,
an intrusion detection system detects
the different type of attacks and sends the report to base
station. It uses all nodes or some special nodes to detect
these types of attacks. These nodes co-operate each other
to take the decision and finally send the report to the base
station. It requires lots of communication between the
nodes. If adversary can trap the message exchanging
between the nodes then he can easily tamper the
messages and send the false information to the other
nodes. Secure communication is a necessary
condition in order to make the network smooth so that
nodes can send data or exchange the message securely. In our paper, we provide the 
definition of a communication system for WSNs, and define some of the required security properties dedicated to sensor networks.
 
{\bf Secure scheduling:} The main objective of a secure scheduling is to prolong the whole network lifetime while fulfilling
the surveillance application needs. In other words, a common
approach is to define a subset of the deployed nodes to be active while the other nodes
can sleep. In this paper, we present a novel scheduling algorithm
where only a subset of nodes contribute significantly to detect intruders and
prevent malicious attacker to predict the behavior of the network prior to intrusion.
We present a random scheduling
to solve this issue, by guaranteeing an uniform coverage
while preventing attackers to predict the list of awaken nodes.

The remainder of this research work is organized as follows.
In the next section we provide a general presentation for security in WSN. A rigorous formalism for secure
communications in wireless sensor networks is presented in~\ref{sec:secure comm}, in which the notions of communication systems, indistinguability, nonmalleability,
and message detection resistance are formalized rigorously in the WSN framework.
In Section~\ref{sc:Secure Scheduling}, the notion of secure scheduling is defined
and applied on a given example. This research work ends by a conclusion section.

%%%%%%%%%%%%%%%%%%%%%%%%%%%%%%%%%%%%%%%%%%%%%%%%%%%%%%%%%%%%%%%%%%%%%%%%%%%%%%%%%%%%%%%%%%%%%%%%%%%%%%%%%%%%%%%%%%%%%%%%%%%%%%%%%%%%%%%%%%%
%%%%%%%%%%%%%%%%%%%%%%%%%%%%%%%%%%%%%%%%%%%%%%%%%%%%%%%%%%%%%%%%%%%%%%%%%%%%%%%%%%%%%%%%%%%%%%%%%%%%%%%%%%%%%%%%%%%%%%%%%%%%%%%%%%%%%%%%%%%%%
%%%%%%%%%%%%%%%%%%%%%%%%%%%%%%%%%%%%%%%%%%%%%%%%%%%%%%%%%%%%%%%%%%%%%%%%%%%%%%%%%%%%%%%%%%%%%%%%%%%%%%%%%%%%%%%%%%%%%%%%%%%%%%%%%%%%%%%%%%%%%

\section{Security in WSN: General presentation}

Wireless nature of communication, lack of
infrastructure and uncontrolled environment improve
capabilities of adversaries in WSN. Stationary adversaries
equipped with powerful computers and communication
devices may access whole WSN from a remote location.
They can gain mobility by using powerful laptops,
batteries and antennas, and move around or within the
WSN. In this section, we consider a WSN where nodes communicate together by sending 
data publicly. These \emph{transmitted data} contain a \emph{message} whose confidentiality 
must be preserved. For instance, transmitted data is the cryptogram of a message, modulated in an electromagnetic radiation, or the message is dissimulated into the electromagnetic radiation by using a spread spectrum information hiding technique.

Wireless communication helps adversaries to
perform variety of attacks. A secure communication can be used to provide the following general 
security goals:

{\bf One-wayness (OW)}, the adversary who sees transmitted data is not able to 
compute the corresponding message.

{\bf Indistinguability (IND)}, observing transmitted data, the adversary learns 
nothing about the contained message.

{\bf Non-malleability (NM)}, the adversary, observing data for a message $m$, 
cannot derive another data for a meaningful message $m'$ related to $m$.

The OW and IND goals relate to the confidentiality of messages sending 
through the WDN. The IND goal is, however, much more difficult to achieve than 
the one-wayness. %Note that probabilistic steganography presented above provides Indistinguability (also termed semantic security).
Non-malleability guarantees that any attempt to manipulate the observed data to 
obtain a valid data will be unsuccessful (with a high probability).

The power of a polynomial attacker (with polynomial computing resources) very much depends on his/her knowledge about the system used to transform \emph{information} in \emph{data}. 
The weakest attacker is an outsider who knows the public embedding algorithm together with other public information about the setup of the system.
The strongest attacker seems to be an insider (he/she is inside the network) who can access the extraction device (recovering information from data) in regular interval.
The access to the extraction key is not possible as the extraction device is assumed to be tamperproof.

An \emph{extraction oracle} is a formalism that mimics an attacker's access to the extraction device.
The attacker can experiment with it providing \emph{data} and collecting corresponding \emph{information} from the oracle (the attacker cannot access to the decryption key).
In general, the public-key WSN can be subjected to the following attacks (ordered in increasing strength):

{\bf Chosen information attack (CIA)} The attacker knows the embedding algorithm and the public elements including the public key (the embedding oracle is publicly accessible).

{\bf Nonadaptative chosen data attack (CDA1)} The attacker has access to the extraction oracle before he sees a data that he wishes to manipulate.

{\bf Adaptative chosen data attack (CDA2)} The attacker has access to the extraction oracle before and after he observes a data $s$ that he wishes to manipulate (assuming that he is not allowed to query the oracle about the data $s$).

\begin{figure}
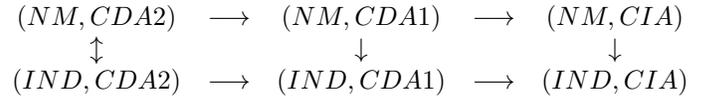

$$
\begin{array}{ccccc}
(NM,CDA2) & \longrightarrow & (NM, CDA1) & \longrightarrow & (NM,CIA)\\
\updownarrow & & \downarrow & & \downarrow\\
(IND,CDA2) & \longrightarrow & (IND, CDA1) & \longrightarrow & (IND,CIA)
\end{array}
$$
\caption{Relations among security notions}
\label{fig:rel}
\end{figure}

The security level that a public-key WSN achieves can be specified by the pair (goal, attack), where the goal can be either OW, IND, or NM, and the attack can be either CIA, CDA1, or CDA2.
For example, the level (NM,CIA) assigned to a public-key network says that the system is nonmalleable under the chosen message attack.
There are two sequences of trivial implications
\begin{itemize}
\item (NM,CDA2) $\Rightarrow$ (NM, CDA1) $\Rightarrow$ (NM,CIA),
\item (IND,CDA2) $\Rightarrow$ (IND, CDA1) $\Rightarrow$ (IND,CIA),
\end{itemize}
which are true because the amount of information available to the attacker in CIA, CDA1, and CDA2 grows.
Figure~\ref{fig:rel} shows the interrelation among different security notions.
Consequently, we can identify the hierarchy of security levels.
The top level is occupied by $(NM,CDA2)$ and $(IND,CDA2)$. 
The bottom level contains $(IND,CIA)$ only as the weakest level of security.
If we are after the strongest security level, its enough to prove that our network attains the $(IND,CDA2)$ level of security.

%%%%%%%%%%%%%%%%%%%%%%%%%%%%%%%%%%%%%%%%%%%%%%%%%%%%%%%%%%%%%%%%%%%%%%%%%%%%%%%%%%%%%%%%%%%%%%%%%%%%%%%%%%%%%%%%%%%%%%%%%%%%%%%%%%%%%%%%%%%
%%%%%%%%%%%%%%%%%%%%%%%%%%%%%%%%%%%%%%%%%%%%%%%%%%%%%%%%%%%%%%%%%%%%%%%%%%%%%%%%%%%%%%%%%%%%%%%%%%%%%%%%%%%%%%%%%%%%%%%%%%%%%%%%%%%%%%%%%%%%%
%%%%%%%%%%%%%%%%%%%%%%%%%%%%%%%%%%%%%%%%%%%%%%%%%%%%%%%%%%%%%%%%%%%%%%%%%%%%%%%%%%%%%%%%%%%%%%%%%%%%%%%%%%%%%%%%%%%%%%%%%%%%%%%%%%%%%%%%%%%%%

\section{Rigorous Formalism for Secure Communications in WSNs}
\label{sec:secure comm}

In this section, we present a new principles formalism for secure communication in wireless sensor networks. 

\subsection{Communication System in a WSN}

\begin{definition}[Communication system]
\label{def:secure system}
Let $\mathcal{S}, \mathcal{M}$, and $\mathcal{K}=\{0,1\}^\ell$ 
three sets of words on $\{0,1\}$ called respectively the sets of 
transmission supports, of messages, and of keys (of size $\ell$).

A \emph{communication system} on $(\mathcal{S}, \mathcal{M}, \mathcal{K})$  
is a tuple $(\mathcal{I},\mathcal{E}, inv)$ such that:
\begin{itemize}
\item $\mathcal{I}:\mathcal{S} \times \mathcal{M} \times \mathcal{K} \longrightarrow \mathcal{S}$, 
$(s,m,k) \longmapsto \mathcal{I}(s,m,k)=s'$, is the \emph{insertion function}, which 
put the message $m$ into the support of transmission $s$ according to the key
$k$, leading to the transmitted data $s'$.
\item $\mathcal{E}:\mathcal{S}  \times \mathcal{K} \longrightarrow \mathcal{M}$, 
$(s,k) \longmapsto \mathcal{E}(s,k) = m'$, defined as the \emph{extraction function}, 
which extract a message $m'$ from a transmitted data $s$, depending on a key $k$.
\item $inv:\mathcal{K} \longrightarrow \mathcal{K}$, s.t. $\forall k \in
  \mathcal{K}, \forall (s,m)\in \mathcal{S}\times\mathcal{M},
  \mathcal{E}(\mathcal{I}(s,m,k),inv(k))=m$, which is the function that can 
``invert'' the effects of the key $k$, producing the message $m$ that has been
embedded into $s$ using $k$.
\item $\mathcal{I}$ and $\mathcal{E}$ can be computed in polynomial time, 
and $\mathcal{I}$ is a probabilistic algorithm (the same  values inputted 
twice produce two different transmitted data).
\end{itemize}
$k$ is called the embedding key and $k'=inv(k)$ the extraction key. If 
$\forall k \in \mathcal{K}, k=inv(k)$, the communication system through the
WSN is said \emph{symmetric} (private-key), otherwise it is \emph{asymmetric}
 (public-key).
\end{definition}

\subsection{Indistinguability}

Suppose that the adversary has two messages $m_1,m_2$ and a transmitted data 
$s$ in his/her possession. He/she knows that $s$ contains either $m_1$ or $m_2$.
Our intention is to define the fact that, having all these materials, the key, 
and the insertion function (we take place into the (IND,CIA) context), he cannot
determine with a non negligible probability the message that has been embedded
into $s$.

The difficulty of the challenge comes, for a large extend, from the fact that
the insertion algorithm $\mathcal{I}$ is a probabilistic one, which is a
common-sense assumption usually required in cryptography.

\begin{definition}
An Indistinguability I-adversary is a couple
$(\mathsf{A}_1,\mathsf{A}_2)$ of nonuniform algorithms, each with access to
an oracle $\mathcal{O}$.
\end{definition}

\begin{definition}[Indistinguability]
For a public communication system in WSN $(\mathcal{I},\mathcal{E}, inv)$ on 
$(\mathcal{S}, \mathcal{M}, \{0,1\}^\ell)$, define the advantage of an 
I-adversary $\mathsf{A}$ by
\begin{tiny}
$$Adv_{\mathsf{A}}^{I-\mathcal{O}} = Pr\left[
\begin{array}{c} k \xleftarrow{\$} \{0,1\}^\ell \\ (m_0,m_1,s) \leftarrow \mathsf{A}_1(k) \\ b \leftarrow \{0,1\} \\ \alpha = \mathcal{I}(s,m_b,k)\end{array}:\begin{array}{c} \mathsf{A}_2(k,s,m_1,m_2,\alpha) = b\end{array}\right]$$
\end{tiny}

\end{definition}

We define the insecurity of $S=(\mathcal{I},\mathcal{E}, inv)$ with respect to 
the Indistinguability as

$$InSec_S^{I-\mathcal{O}}(t) = \max_\mathsf{A}\left\{Adv_\mathsf{A}^{I-\mathcal{O}}\right\}$$
where the maximum is taken over all adversaries $\mathsf{A}$ with total running time $t$.

We distinguish three kinds of oracles:
\begin{itemize}
\item The Non-adaptative oracle, denoted $\mathcal{NA}$, where $A_1$ and
  $A_2$ can only access to the elements of the communication system. 
\item The Adaptative oracle, denoted $\mathcal{AD}1$, where $A_1$ has access the
  communication system and to an oracle that can in a constant time provide a 
message  $m^\prime$ from any transmitted data
  $\mathcal{I}(M^\prime,m^\prime,k^\prime)$, without knowing neither $M^\prime$ nor
  $k^\prime$ nor $inv(k^\prime)$. In this context, $A_2$ has no 
  access to this oracle.
\item The Strong adaptative oracle, denoted $\mathcal{AD}2$, where $A_1$ has access to the
  communication system and to an oracle that can in a constant time provide a message
  $m^\prime$ from any transmitted data
  $\mathcal{I}(M^\prime,m^\prime,k^\prime)$, without knowing neither $M$ nor
  $k^\prime$ nor $inv(k^\prime)$. In this context, $A_2$ has also access to
  this oracle but for the message $\mathcal{I}(M,m_b,k)$.
\end{itemize}

\subsection{Relation Based Non-malleability}

In some scenarios malicious nodes can integrate the WSN, hoping by doing so
to communicate false information to the other nodes. We naturally suppose that
communications are secured. The problem can be formulated as follows: is it 
possible for the attacker to take benefits from his/her observations, in order
to forge transmitted data either by embedding erroneous messages, or sending data that appear to be similar
with what a node is supposed to produce.

As wireless sensor networks have usually a dynamical architecture, the 
(dis)appearance of nodes is not necessarily suspect. Authentication protocols
can be deployed into the WSN, but in some cases such authentication is 
irrelevant, because of its energy consumption, communication cost, or rigidity.
We focus in this section on the possibility to propose a secured communication
scheme in WSN that prevents an attacker to forge such malicious transmitted
data. Such non-malleability property can be formulated as follows.

\begin{definition}
A Relation Based NM-adversary is a  
nonuniform  algorithm $\mathsf{A}$ having access to an oracle $\mathcal{O}$.
\end{definition}

\begin{definition}[Relation Based Non-malleability]
For a public communication system  $(\mathcal{I},\mathcal{E}, inv)$ on 
$(\mathcal{S}, \mathcal{M}, \{0,1\}^\ell)$, define the advantage of a 
NM-adversary $\mathsf{A}$ by
\begin{tiny}
$$~~~~~~Adv_{\mathsf{A}}^{NM-\mathcal{O}}(m) = Pr\left[
\begin{array}{c} 
s \leftarrow \mathcal{S} \\ 
k \xleftarrow{\$} \{0,1\}^\ell \\ 
s' \leftarrow \mathsf{A}(\mathcal{I}(s,m,k)) \\
m' \leftarrow  \mathcal{E}(s',k)
\end{array}
:\begin{array}{c} m' \in R(m) \end{array}\right]$$
\end{tiny}
\end{definition}
where $R:\mathcal{M} \longrightarrow \mathcal{P}(\mathcal{M})$ is a function that
map any message $m$ to a subset of $\mathcal{M}$ containing messages related
to $m$ (for a given property).
For instance, if we suppose that an attacker has inserted or corrupted some nodes
in a network that measures temperature, he can make these nodes 
send wrong temperatures values fixed {\it a priori}.
%If the targeted temperature is 20\textdegree C, the attacker can accept, when trying to
%forge the transmitted data, temperatures that range into $R(20)=\{19, 20, 21\}$,
 %due to the difficulty of the task.

We can now define the insecurity of $S=(\mathcal{I},\mathcal{E}, inv)$ with respect to 
the Relation Based Non-malleability as
$$InSec_S^{NM-\mathcal{O}}(t) = \max_\mathsf{A}\left\{\max_{m \in \mathcal{M}}\left\{Adv_\mathsf{A}^{NM-\mathcal{O}}(m)\right\}\right\}$$
where the maximum is taken over all adversaries $\mathsf{A}$ with total running 
time $t$.
Similar kinds of oracles than previously can be defined in that context.

\subsection{Message Detection Resistance}

We now address the particular case where transmitted data can contain or not
an embedded message. For security reasons, it is sometimes required that an
attacker cannot determine when information are transmitted through the network.
For instance, in a video surveillance context, suppose that an attacker can 
determine when an intrusion is detected, or when something considered as 
suspicious is forwarded through the nodes to the sink. Then he/she can use this
knowledge to deduce what kind of behavior is suspicious for the network, adapting
so his/her attacks. Decoys are often proposed to make such attacks impossible: 
transmitted data do not always contain information, some of the communications
are only realized to mislead the attacker. The quantity and frequency of these
decoys must naturally respect the energy consumption requirement, and a 
compromise must be found on the message/decoy rate to face such attacks while
preserving the WSN lifetime. However, such an approach supposes that the attacker
is unable to make the distinction between decoys and meaningful communications.
Such a supposition leads to the following definition.

\begin{definition}
A Detection Resistance DR-adversary is a couple $(\mathsf{A}_1,\mathsf{A}_2)$ of 
nonuniform  algorithms, each with access to an oracle $\mathcal{O}$.
\end{definition}

\begin{definition}[Message Detection Resistance]
For a public communication system  $(\mathcal{I},\mathcal{E}, inv)$ on 
$(\mathcal{S}, \mathcal{M}, \{0,1\}^\ell)$, define the advantage of a 
DR-adversary $\mathsf{A}$ by
\begin{tiny}
$$\begin{small}Adv_{\mathsf{A}}^{DR-\mathcal{O}} = Pr\left[
\begin{array}{c} M_0, M_1 \leftarrow \mathcal{S} \\ k \xleftarrow{\$} \{0,1\}^\ell \\ m \leftarrow \mathsf{A}_1(k) \\ b \leftarrow \{0,1\} \\ \alpha = \left\{M_b, \mathcal{I}(M_{\overline{b}},m,k))\right\}\end{array}:\begin{array}{c} \mathsf{A}_2(m,k,\alpha) = M_b\end{array}\right]\end{small}$$
\end{tiny}
\end{definition}

\noindent where the set defining $\alpha$ is a non-ordered one.

We define the insecurity of $S=(\mathcal{I},\mathcal{E}, inv)$ with respect to 
the Message Detection Resistance as
$$InSec_S^{DR-\mathcal{O}}(t) = \max_\mathsf{A}\left\{Adv_\mathsf{A}^{DR-\mathcal{O}}\right\}$$
where the maximum is taken over all adversaries $\mathsf{A}$ with total running 
time $t$.
Similar kinds of oracles than previously can be defined in that context.

%%%%%%%%%%%%%%%%%%%%%%%%%%%%%%%%%%%%%%%%%%%%%%%%%%%%%%%%%%%%%%%%%%%%%%%%%%%%%%%%%%%%%%%%%%%%%%%%%%%%%%%%%%%%%%%%%%%%%%%%%%%%%%%%%%%%%%%%%%%
%%%%%%%%%%%%%%%%%%%%%%%%%%%%%%%%%%%%%%%%%%%%%%%%%%%%%%%%%%%%%%%%%%%%%%%%%%%%%%%%%%%%%%%%%%%%%%%%%%%%%%%%%%%%%%%%%%%%%%%%%%%%%%%%%%%%%%%%%%%%%
%%%%%%%%%%%%%%%%%%%%%%%%%%%%%%%%%%%%%%%%%%%%%%%%%%%%%%%%%%%%%%%%%%%%%%%%%%%%%%%%%%%%%%%%%%%%%%%%%%%%%%%%%%%%%%%%%%%%%%%%%%%%%%%%%%%%%%%%%%%%%

\section{Secure Scheduling}
\label{sc:Secure Scheduling}
\subsection{Motivations}

A common way to enlarge lifetime of a wireless sensor network is
to consider that not all of the nodes have to be awakened: a subset
of well-chosen nodes participates temporarily to the task devoted 
to the network~\cite{91}, whereas the other nodes sleep in order to
preserve their batteries. Obviously, the scheduling process 
determining the nodes that have to be awakened at each time must
be defined carefully, both for guaranteeing a certain level of
quality in the assigned task and to preserve the network capability
over time. Problems that are of importance in that approach are often
related to coverage, ratio of awaken vs sleeping nodes, efficient transmission of wake up orders, and capability for the partial network
to satisfy, with a sufficient quality, the objectives it has been
designed for. Existing surveillance applications works focus on finding an efficient deployment pattern so that the average overlapping area of each sensor is bounded. 
The authors in~\cite{94} analyze new deployment strategies for satisfying some given coverage probability requirements with directional sensing models. 
A model of directed communications is introduced to ensure and repair the network connectivity.
Based on a rotatable directional sensing model, the authors in~\cite{95} present a method to deterministically estimate the amount of directional nodes for a given coverage rate. A sensing connected sub-graph accompanied with a convex hull method is introduced to model a directional sensor network into several parts in a distributed manner. 
With adjustable sensing directions, the coverage algorithm tries to minimize the overlapping sensing area of directional sensors only with local topology information.
Lastly, in~\cite{93}, the authors present a distributed algorithm that ensures both coverage of the deployment area and network connectivity, by providing multiple cover sets to manage Field of View redundancies and reduce objects disambiguation.

All the above algorithms depend on the geographical location information of sensor nodes. 
These algorithms aim to provide a complete-coverage network so that any point in the target area would be covered by at least one sensor node. 
However, this strategy is not as energy-efficient as what we expect because of the following two reasons.
Firstly, the energy cost and system complexity involved in obtaining geometric information may compromise the effect of those algorithms. 
Secondly, sensor nodes located at the edge of the area of interest must be always in an active state as long as the region is required to be completely covered. 
These nodes will die after some time and their coverage area will be left without surveillance. 
Thus, the network coverage area will shrink gradually from outside to inside. 
This condition is unacceptable in surveillance applications and (intelligent) intrusion detection, because the major goal here is to detect intruders as they cross a border or as they penetrate a protected area. In case of hostile environments, security play an 
important role in the written of the scheduling program. Indeed 
an attacker, observing the manner nodes are waken up, should not
be able to determine the scheduling program. For instance, in a 
video surveillance context, if the attacker is able to determine at
some time the list of the sleeping nodes, then he can possibly achieve
an intrusion without being detected~\cite{bgmp11:ip}. 

Obviously, a random scheduling
can solve the issues raised above, by guaranteeing a uniform coverage
while preventing attackers to predict the list of awaken nodes. 
However, this approach needs random generators into nodes, which 
cannot be obtained by deterministic algorithms embedded into the
network. Even if truly random generators (TRG) can be approximated by
physical devices, they need a certain quantity of resources, suppose
that the environment under observation has a sufficient variability
of a given set of physical properties (to produce the physical
noise source required in that TRG), and are less flexible or adaptable
on demand than pseudorandom number generators (PRNGs). Furthermore, neither
their randomness nor their security can be mathematically proven: these generators
can be biased or wrongly designed. Being able to guarantee a certain level of security in scheduling
leads to the notion of \emph{secure scheduling} proposed below.

\subsection{Secure Scheduling in Wireless Sensor Networks}

Two kinds of scheduling processes can be defined: each node can embed its own
program, determining when it has to sleep (local approach), or
the sink or some specific nodes can be responsible of the scheduling
process, sending sleep or wake up orders to the nodes that have to
change their states (global approach). 

We consider that a deterministic scheduling algorithm is a function $S: \{0,1\}^n \rightarrow \{0,1\}^M$, where $M > n$. This definition can be understood as follows:
\begin{itemize}
\item The value inputted in $S$ is the secret key launching the scheduling process. It can be shown as the seed of a PRNG.
\item In case of a local approach, the binary sequence produced by this
function corresponds to the moments where the node must be awaken: if
the $k-$th term of this sequence is 0, then the node can go to sleep mode between
$t_k$ and $t_{k+1}$.
\item In case of a global approach, the binary sequence returned by $S$ can be divided by blocs, such that each bloc contains the id of the node to which an order of state change will be send.
\end{itemize}

Loosely speaking, $S$ is called a secure scheduling if it maps uniformly distributed input (the secret key or seed of the scheduling process) into an output which is computationally indistinguishable from uniform. 
The precise definition is given below.

\begin{definition}
A $T$-time algorithm $\mathcal{D} : \{0, 1\}^M \longrightarrow {0, 1}$ is said to be a $(T,\varepsilon)$-distinguisher for $S$ if
$$\left| Pr[\mathcal{D}(S(\mathfrak{U}_2^n )) = 1] - Pr[\mathcal{D}(\mathfrak{U}_2^M) = 1] \right| \geqslant \varepsilon.$$
\end{definition}

\noindent where $\mathfrak{U}_2$ is the uniform distribution on $\{0,1\}$.

\begin{definition}[Secure scheduling]
Algorithm $S$ is called a $(T,\varepsilon )$-secure scheduling 
if no $(T,\varepsilon)$-distinguisher exists for $S$.
\end{definition}

Adapting the proofs of~\cite{Yao82, Goldreich86}, it is possible to show that 
a $(T,\varepsilon)-$distinguisher exists if and only if a $T$-time algorithm
can, knowing the first
$l$ bits of a scheduling $s$, predict the $(l + 1)-$st bit of $s$ with probability
significantly greater than $0.5$.
This comes from the fact that a PRNG passes the next-bit test if and only if it
 passes all polynomial-time statistical tests~\cite{Yao82, Goldreich86}.

An important question is what level of security $(T,\varepsilon)$ suffices for practical applications
in scheduled wireless sensor networks. 
Unfortunately, the level of security is often chosen arbitrarily.
It is reasonable to require that a scheduling process is secure for all pairs $(T,\varepsilon)$
such that the time-success ratio $T/\varepsilon$ is below a certain bound. In the next section we present an illustration of this notion.

%Knuth ([14], p. 176) sets = 0.01 and consider several values for T up to 53.5 · 1012 Mips-
%Years1 . In [6], the security level is set to T = 1 Mips-Year and = 0.01. In [8], T = 3.5 · 1010
%%Mips-Years, = 0.01.
%The fact that a pseudorandom generator is (T, )-secure does not automatically mean
%that the generator is (T , )-secure for all T and such that T / ≤ T / . For instance, if a
%pseudorandom generator is (T, 0.01)-secure it does not necessarily mean that the generator
%is (T , 0.009)-secure even if T
%T . The reason is that a (T , 0.009)-distinguisher cannot
%always be transformed into a (T, 0.01)-distinguisher. Indeed, the only way to improve the
%success probability of the distinguisher is to run it several times on the same input. However,
%the latter does not always help since there might be ”bad” inputs, that is, inputs for which
%the success probability of the distinguisher is very low or equals 0.
%that is set to be 280 time
%units throughout this paper (the time unit is defined in Section 2.4). Time-success ratio is
%a standard way to define security of cryptographic schemes [17, 27].
%1
%A Mips-Year is defined as the amount of computation that can be performed in one year by a
%single DEC VAX 11/780 (see also [16]).

\subsection{Practical Study}

Suppose that a wireless sensor node has been 
scheduled by a Blum-Blum-Shub BBS pseudorandom
generator. This generator produces bits 
$y_0, y_1, \hdots$, and the node is awaken during
the time interval $[t_i;t_{i+1}[$ if and only if
 $y_i=1$.

Let us recall that the Blum Blum Shum 
generator~\cite{Blum:1985:EPP:19478.19501} (usually denoted by BBS) is defined
by the following process:
\begin{enumerate}
\item Generate two large secret random and distinct primes $p$ and $q$, each congruent to 3 modulo 4, and compute $N = pq$.
\item Select a random and secret seed $s \in \llbracket 1, N - 1 \rrbracket$ such that $gcd(s, N) = 1$, and compute $x_0 = s^2 (\textrm{mod } N)$.
\item For $1 \leqslant i \leqslant l$ do the following:
\begin{enumerate}
\item $x_i = x_{i-1}^2 (\textrm{mod } N)$.
\item $y_i =$ the least significant bit of $x_i$.
\end{enumerate}
\item The output sequence is $y_1 , y_2 , \hdots, y_l$.
\end{enumerate}

Suppose now that the network will work during 
$M=100$ time units, and that during this period,
an attacker can realize $10^{12}$ clock cycles.
We thus wonder whether, during the network's 
lifetime, the attacker can distinguish this 
sequence from truly random one, with a probability
greater than $\varepsilon = 0.2$.
We consider that $N$ has 900 bits.

The scheduling process is the BBS generator, which
is cryptographically secure. More precisely, it
is $(T,\varepsilon)-$secure: no 
$(T,\varepsilon)-$distinguishing attack can be
successfully realized on this PRNG, if~\cite{Fischlin}
$$
T \leqslant \dfrac{L(N)}{6 N (log_2(N))\varepsilon^{-2}M^2}-2^7 N \varepsilon^{-2} M^2 log_2 (8 N \varepsilon^{-1}M)
$$
where $M$ is the length of the output ($M=100$ in
our example), and
$$
L(N)=2.8\times 10^{-3} exp \left(1.9229 \times (N ~ln(2)^\frac{1}{3}) \times ln(N~ln 2)^\frac{2}{3}\right)
$$
is the number of clock cycles to factor a $N-$bit
integer.

A direct numerical application shows that this attacker 
cannot achieve its $(10^{12},0.2)$ distinguishing
attack in that context.

\subsection{Results study}

\begin{figure}[h]  
\centering  
\includegraphics[width=\linewidth]{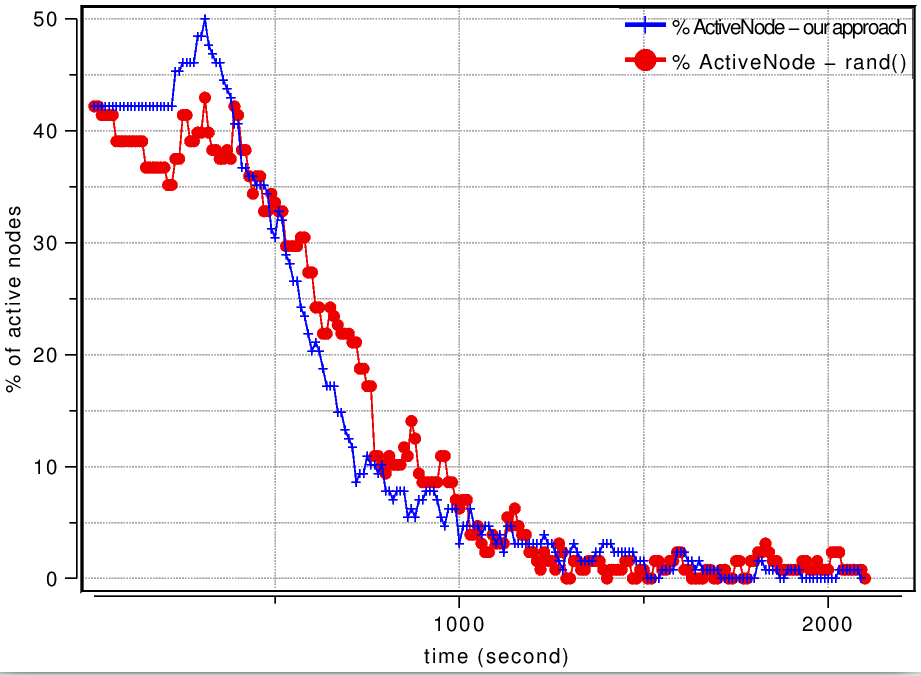}  
%\vspace{-1mm}  
\caption{Percentage of active nodes.}   
\label{graph-activenode}  
\end{figure}

This section presents simulation results on comparing our approach to the standard C++ {\tt rand()}-based approach with random intrusions. We use the OMNET++ simulation environment and the next node selection will either use our approach or the C++ {\tt rand()} function ({\tt rand() \% $2^n$}) to produce a random number between 0 and $2^n$. For these set of simulations, $128$ sensor nodes are randomly deployed in a $75m * 75m$ area. Figure \ref{graph-activenode} shows the percentage of active nodes. Both our approach and the standard {\tt rand()} function have similar behavior: the percentage of active nodes progressively decreases due to battery shortage.

\begin{figure}[h]  
\centering  
\includegraphics[width=\linewidth]{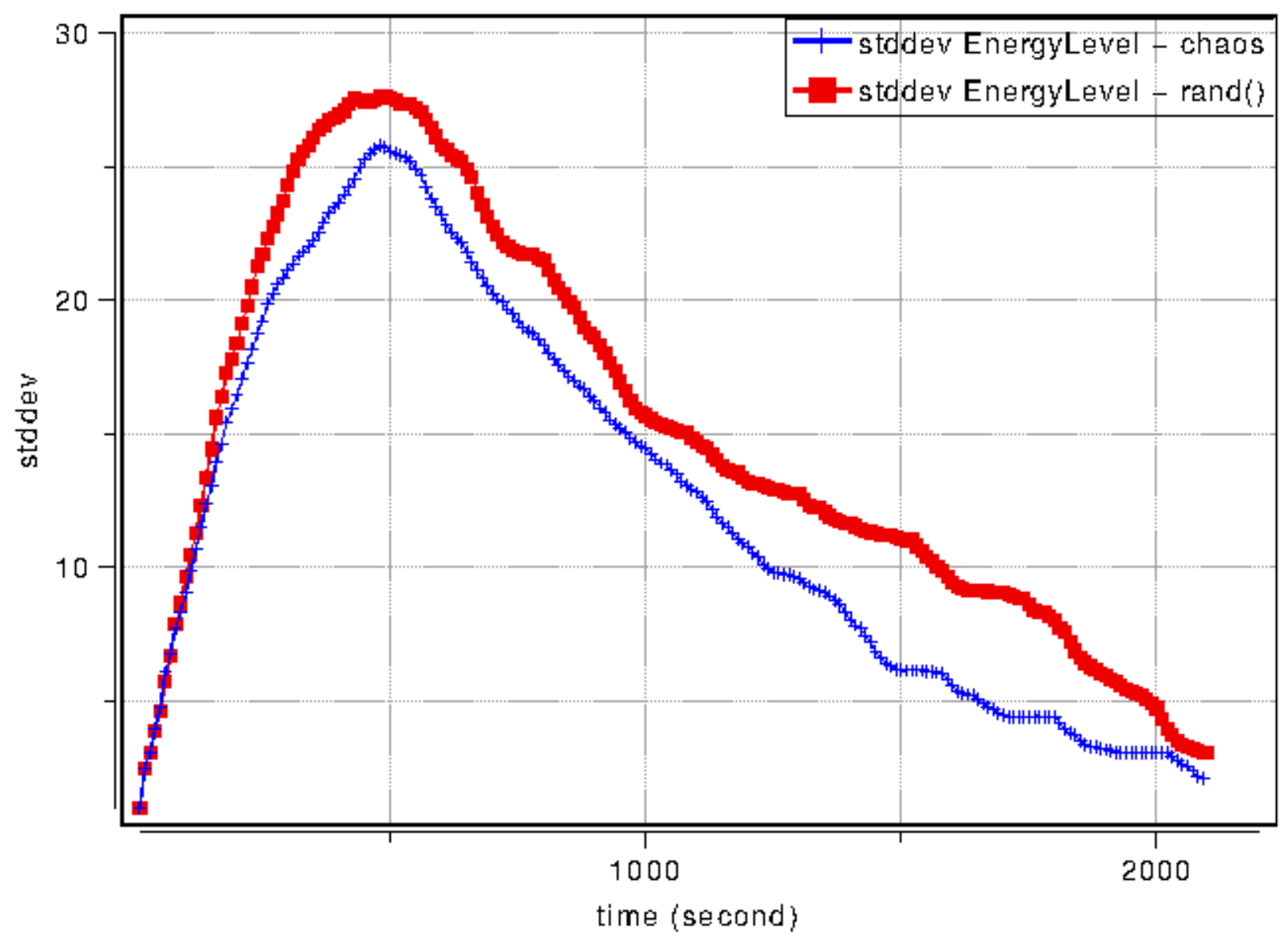}  
%\vspace{-1mm}  
\caption{Evolution of the energy consumption's standard deviation.}   
\label{graph-stddev}  
\end{figure}

Another result we want to show is the energy consumption distribution. We recorded every 10s the energy level of each sensor node in the field and computed the mean and the standard deviation. Figure \ref{graph-stddev} shows the evolution of the standard deviation during the network lifetime. We can see that our approach selection provides a slightly better distribution of activity than the standard {\tt rand()} function.

%%%%%%%%%%%%%%%%%%%%%%%%%%%%%%%%%%%%%%%%%%%%%%%%%%%%%%%%%%%%%%%%%%%%%%%%%%%%%%%%%%%%%%%%%%%%%%%%%%%%%%%%%%%%%%%%%%%%%%%%%%%%%%%%%%%%%%%%%%%
%%%%%%%%%%%%%%%%%%%%%%%%%%%%%%%%%%%%%%%%%%%%%%%%%%%%%%%%%%%%%%%%%%%%%%%%%%%%%%%%%%%%%%%%%%%%%%%%%%%%%%%%%%%%%%%%%%%%%%%%%%%%%%%%%%%%%%%%%%%%%
%%%%%%%%%%%%%%%%%%%%%%%%%%%%%%%%%%%%%%%%%%%%%%%%%%%%%%%%%%%%%%%%%%%%%%%%%%%%%%%%%%%%%%%%%%%%%%%%%%%%%%%%%%%%%%%%%%%%%%%%%%%%%%%%%%%%%%%%%%%%%

\section{Conclusion}

In this document, a rigorous framework for security in
wireless sensor networks has been formalized. The 
definition of a communication system in WSNs has been
introduced, and security properties (indistinguability,
nonmalleability, and message detection resistance) have
been formalized in that context. Furthermore, the
definitions of secure scheduling, specific to such networks, have been
given too. With this theoretical framework, it has been
possible to evaluate the security of a scheduling scheme
based on the BBS cryptographically secure PRNG.

\bibliographystyle{unsrt}
\bibliography{mabase}

\end{document}